# Delayed neutrons measurement at the MEGAPIE target

S. Panebianco[1,a], P. Bokov[1], D. Doré[1], X. Ledoux[2], A. Letourneau[1], A. Prévost[1], D. Ridikas[1]

[1] CEA Saclay, DSM/DAPNIA/SPhN, F-91191 Gif-sur-Yvette Cedex, France
[2] CEA DIF, DAM/DPTA/SPN, F-91680 Bruyères-le-Châtel, France

**Abstract.** In the framework of the Neutronic and Nuclear Assessment Task Group of the MEGAPIE experiment we measured the delayed neutron (DN) flux at the top of the target. The measurement was proposed mainly for radioprotection purposes since the DN flux at the top of the target has been estimated to be of the same order of magnitude as the prompt neutron flux. Given the strong model-dependence of DN predictions, the measurement of DN contribution to the total neutron activity at the top of the target was thus desired. Moreover, this measurement is complementary to the DN experiments performed at PNPI (Gatchina) on solid lead and bismuth targets. The DN measurement at MEGAPIE was performed during the start-up phase of the target. In this paper we present a detailed description of the experimental setup and some preliminary results on decay spectra.

## 1 Introduction

In the framework of the R&D on future spallation sources, neutrino factories or RIB facilities, radioprotection issues are clearly of major concern. The increased proton beam power needed by new generation targets results in heat deposition constraints leading to innovative designs based on heavy liquid metal (HLM). Present spallation targets aiming at beam power of 1 MW or higher (e.g., MEGAPIE (PSI), SNS (US), JSNS (Japan), ESS and EURISOL (both Europe)), all focus on liquid metal targets.

In fact the use of liquid metal loop can solve some difficult problems for high-power spallation sources, mainly related to the evacuation of the deposited heat. On the other hand, it also introduces some new issues that must be addressed (e.g., corrosion, resistance of the target window or interface accelerator, etc.). In addition, radioactive nuclides produced in liquid metal targets are transported into hot cells, pumps or close to electronics with radiation sensitive components. Besides the considerable amount of decay gamma activity in the irradiated liquid metal, a significant amount of the Delayed Neutron (DN) precursor activity can be accumulated in the target fluid. The transit time from the front of a liquid metal target into areas where DNs may be important, can be as short as a few seconds, i.e. well within one half-life of many DN precursors. Therefore, it seems very important to evaluate the DN flux as a function of position and determine if DNs may contribute significantly to the activation and dose rates.

In this work we report on the measurement of the DN flux performed during the operation of the MEGAPIE target and we compare it to a simplified geometrical model developed to provide quantitative estimates of the neutron fluxes due to both prompt neutrons (PN) and DNs.

### 1.1 The case of the MEGAPIE target

The realization of spallation neutron sources with nominal beam power of the order of 1 MW or higher started with the SINQ facility (PSI, Switzerland) in year 2000. In this context, MEGAPIE is the first experiment demonstrating the operational feasibility of a spallation target based on liquid metal technology.

MEGAPIE is an international joint initiative to design, build and explore a liquid lead-bismuth spallation target at a beam power of 1 MW [1]. The target has been successfully irradiated during 4 months starting in the middle of August. A big effort was put, since the beginning of the project, to provide the maximum of neutronic information in order to design, operate and qualify such a complex system. This has been the main task of the Neutronic and Nuclear Assessment group (so called X9), which concentrated its neutronic activity on prompt neutrons [2-3].

Since 2003, different authors started calculations of the DNs flux in HLM based systems [4-5], where the importance of this issue for design and radioprotection was outlined. Moreover, it was pointed out that final estimates on DNs are very much model-dependent and no experimental data were available for DN yields from high energy fission-spallation reactions on Pb and Bi targets. This initiated an experimental campaign devoted to characterize the DN production in high energy fission-spallation reactions with Pb and Bi targets [6] [11]. As a natural conclusion of this campaign, the measurement of DNs flux at the MEGAPIE target was proposed to validate the estimations made by Monte Carlo simulation.

---

[a] Presenting author, e-mail: stefano.panebianco@cea.fr

## 2 Estimation of DNs flux at MEGAPIE

Liquid Pb-Bi eutectics (LBE) loop in the case of the MEGAPIE spallation target, as in most of the high power spallation targets based on liquid metal technologies, extends much further compared to the primary proton interaction zone. As it is presented in fig. 1, the activated LBE reaches as high as 400 cm arriving in the heat exchanger, from where it returns to its initial position. It takes ~20 s for the entire ~82 liters of Pb-Bi to make a "round trip" at a flow rate of ~4 liters/s. It is clear that a big part of the DN precursors, created in the interaction region via high energy fission-spallation, will not have enough time to decay completely even at the very top location of the circulating liquid metal. The main concern is about the DNs flux contributing to the total neutron flux at the very top position of the heat exchanger.

### 2.1 A simplified geometrical model

To estimate the DN flux we employed the multi-particle transport code MCNPX [7] combined with the material evolution program CINDER'90 [8], as detailed in [5]. The DN data (emission probabilities and decay constants) were based on the ENDF/B-VI evaluations [9]. For the MEGAPIE target characteristics we used the design values, i.e. a 575 MeV proton beam with 1.75 mA intensity, interacting with the liquid LBE target. The 3-D geometry of the target has been modeled in detail by taking into account all materials used in the design, as described in [10] and [3].
The estimation of the DN parameters for MEGAPIE was performed in steps according to the following procedure:

- calculation of independent fission fragment and spallation product distributions with MCNPX;
- calculation of cumulative fission fragment and spallation product yields with CINDER'90;
- identification of all known DN precursors and construction of the 6-group DN table.

After having built the DN table we developed a generalized geometrical model to estimate the DN activity densities at any position $x$ of the MEGAPIE target loop as presented in fig 1. From the figure we can notice that the LBE cross section changes over the loop, meaning that the transit time of a given LBE volume depends on its position. In particular, the permanence time of an LBE volume under irradiation (in the so called spallation region) is very short (~0.5 s) compared to the total circulation time (~20 s). Within this model the DN activity at position $x$ can be expressed as:

$$a(x) = \sum_{i=1}^{n} a_i(x) = \sum_{i=1}^{n} a_i^a \frac{1-\exp(-\lambda_i \tau_a)}{1-\exp(-\lambda_i T)} \exp(-\lambda_i \tau_d(x)), \quad (1)$$

where $\tau_a$ is the activation time of the Pb-Bi under irradiation; $T$ represents the total circulation period of the LBE, i.e. duration of the "round trip"; $\tau_d$ is the transit (decay) time to reach the point $x$; $\lambda_i$ are the decay constant of the DN precursor $i$ while $a_i$ express the density of DNs due to the precursor $i$.

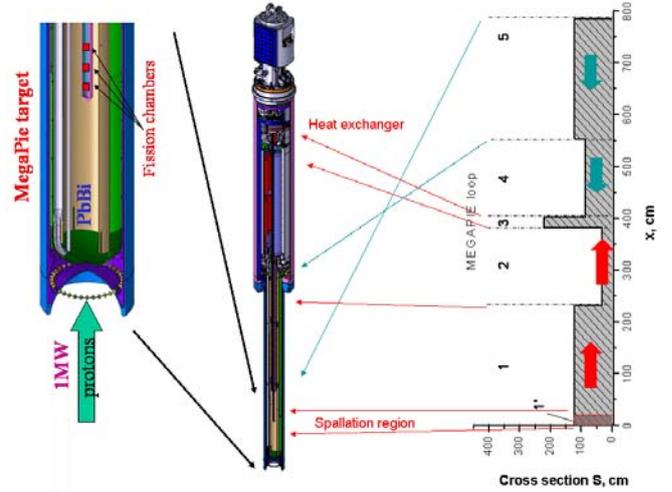

**Fig. 1.** Schematic view of the entire MEGAPIE target (in the middle) with a zoom of the lowest part – proton - Pb-Bi interaction zone (on the left). On the right: cross section of the liquid LBE loop as a function of the Pb-Bi geometrical position – trajectory $x$ (cm)

By the use of the above equation and the 6-group DN table [5] we found that at the very top position of the LBE loop (400 cm level above the target window) the DN activity is of the order of $2 \cdot 10^5$ n/(s cm$^3$). This intermediate result permitted to recalculate the neutron flux at the level of the heat exchanger inserting the volumetric DN source as a function of $x$ provided in fig. 1. It was found that the neutron flux at this position due to DNs and prompt spallation neutrons is of the same order of magnitude, both equal to a few $10^6$ n/(s cm$^2$). It should be pointed out that this estimation rely on the hypothesis that 3 averaged time parameters are sufficient to describe a simplified liquid metal loop dynamics. These time constants, estimated from the target characteristics (LBE volume and main pump speed) are: $\tau_a=0.5s$, $T=20s$ and $\tau_d$ (at the heat exchanger)$=10s$ (see equation 1).

In addition, prompt neutron energy spectrum at the heat exchanger position is very close to thermal (because the MEGAPIE spallation target is surrounded by a heavy water moderator-reflector) while the DN energy spectrum at this level is not "perturbed" yet, i.e. with an average energy of the order of 400-600 keV. These fast neutrons will have considerably higher penetration power compared to the thermal ones. This result clearly points out that activation and dose rates due to DNs should not be neglected.

On the other hand, the 6-group DN parameters, i. e. yields and time spectra of DNs, which were extracted from MCNPX simulations based on different physics models (namely INCL4+ABLA and CEM2k), are model-dependent nearly by two orders of magnitude [5]. This analysis showed that DN yields and time spectra from high energy fission-spallation reactions needed to be measured since no data of this type were available.

## 3 Measurement of DNs at MEGAPIE

In order to verify the estimations of DN flux at MEGAPIE we proposed to measure the DN flux at the top of the target head. Although the inner rod of MEGAPIE target was instrumented with an innovative neutron detector to measure precisely the PN flux [3], the DN flux at this position was too small to obtain with reasonable sensitivity. Thus we decided to make use of a simple setup, based on a $^3$He counter.

### 3.1 Experimental setup and method

The neutron detection was performed by a 45 cm long $^3$He (8 bar) tube counter installed in a polyethylene box (45x20x10 cm$^3$). The $CH_2$ box ensured the moderation of neutrons in order to increase the neutron detector efficiency. The $CH_2$ box was surrounded by a 1 mm thick $^{nat}$Cd foil to avoid the background due to the thermal neutrons. The detector was placed in the target head enclosure chamber, the so called TKE, at around 3 m from the target head.

The detector was set up and tested in the TKE at the end of June 2006. During this phase we performed a complete characterization of the detection system by using an Am-Be neutron source placed in different positions on the polyethylene box. The results of these tests have been compared to a simple MCNP simulation of the detector in order to estimate the neutron detection efficiency. This simulation, which did not take into account the whole geometry of the TKE and was performed by taking a point-like source, showed a good agreement between the measured counting rate and the calculation. In order to estimate the detector efficiency during the experiment we will perform a complete MC simulation of the detector and its environment, taking into account that the delayed neutrons comes from the whole LBE loop

The DN data taking took place during the first week of the target irradiation. During this start-up phase, the beam power was increased in long steps, each step being followed by a beam stop. This procedure gave us the possibility to acquire data starting at each beam stop, corresponding to different beam powers.

The measured counting rate as a function of time, normalized to the beam intensity, is presented on fig 2 for different beam powers. We can notice that, due to the high neutron flux, during the irradiation the neutron detector saturates, i.e. it gives a counting rate independent from the beam power. When the beam stops, we start counting DNs but some seconds are needed before the counting rate becomes well proportional to the beam intensity. Since we know that the detector electronics needs 50 ms after saturation to become operation again, we can argue that the DN flux during the first seconds after the beam stop is still quite high (from calculations on the same order of PN flux) and the detector is still in saturation. As soon as the DN flux lowers, the counting rate becomes well proportional to the beam intensity: as expected, at equilibrium the DN precursor production rates are proportional to the beam power. Thanks to this proportionality, we can sum the decay curves taken at different beam power to increase the statistics. The interpretation of the data is not trivial since we do not know which precursors are contributing to the DN flux. On the other hand, DNs were measured on solid lead and bismuth targets during an experiment at PNPI (Gatchina). This measurement can be a good starting point for the MEGAPIE data interpretation.

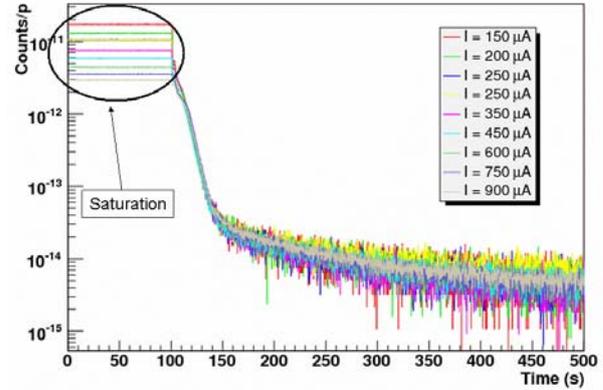

**Fig. 2.** The DN decay curve normalized to the beam intensity; different curves represent different beam powers.

## 4 The experiment at PNPI-Gatchina

In order to constrain the physics models within the MCNPX code, a set of DN measurements was performed using 1 GeV protons interacting with massive Pb and Bi targets of variable thicknesses at PNPI Gatchina (Russia) [11]. After the beam switched off, DNs were detected with an optimized $^3$He detector following specific irradiation periods. Long, intermediate and short irradiation cycles were used to optimize the extraction of different time parameters of DN groups. The $^3$He tube, located at 90° with respect to be beam axis, was surrounded by cylindrical polyethylene ($CH_2$) moderator, surrounded by 1mm $^{nat}$Cd foil. The detector was also shielded by 8-16 cm thick wall made of borated polyethylene.

The analysis of the accumulated data was performed by fitting the decay curves with exponential sums, as shown in fig. 3. It was found that, contrary to the conventional 6-group approach, 4 exponential terms are sufficient to well reproduce the DN decay curves.

Moreover, the major DN contribution, at least up to 10-20 s, comes from light mass products as $^9$Li and $^{17}$N rather than from spallation/fission products as in the case of actinides (see table 1). For longer decay time, from 50 to 100 s, the DN activity is dominated by usual fission products as $^{88}$Br and $^{87}$Br. Finally, the measured absolute yields as a function of the target thickness show, as expected, a similar dependence for the two materials. Moreover, a saturation effect around 20 cm is observed, which means that the production cross sections of DN precursors by fission or fragmentation/evaporation is much more important at higher energies, i.e. the most of the reactions take place in the first half of the stopping target.

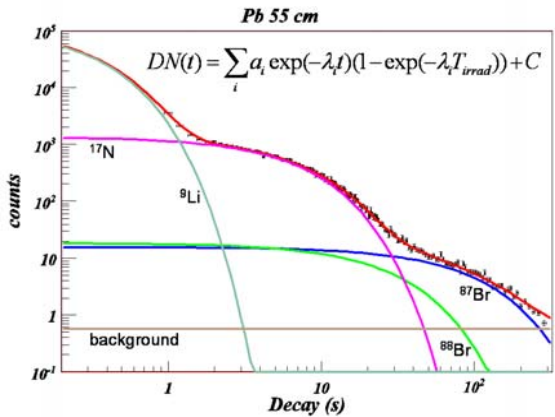

**Fig. 3.** The experimental DN decay curve of a 55 cm thick Pb target, together with the exponential sum fit. Individual group contributions are also indicated.

**Table 1.** Decay parameters of 4 major contributors to the DN yields from high energy fission-spallation reactions.

| Group | Half-life, s | Precursor | Pn (β-n), % |
|---|---|---|---|
| 1 | 55.60 | $^{87}$Br | 2.52 |
| 2 | 16.29 | $^{88}$Br | 6.58 |
| 3 | 4.173 | $^{17}$N | 95.10 |
| 4 | 0.178 | $^{9}$Li | 50.80 |

## 5 Discussion of MEGAPIE results

The analysis of the accumulated data is not yet finalised, thus we present here only a preliminary discussion of the results. As explained above, since the counting rate is proportional to the beam power, the normalized decay spectra shown in fig. 2 can be summed up to increase the statistics. In fig. 4 we show the full statistics, normalized to unity at $t=0$. We can now compare the measured decay curve to the geometrical model presented above. Taking the DN precursor parameters measured at PNPI (see table 1 and Ref. [11]), the LBE transit times estimated from the geometrical model (fig. 1) and inserting them into eq. 1 we can see in the figure that the data (in red) are rather well reproduced by the calculation (in black). This means that in the LBE loop the precursors involved are the same as in the solid target experiment performed at Gatchina. Moreover, the loop thermal hydraulics can be well approximated by the three averaged time parameters. This validates the eq. 1 and the simplified approach developed in [5].

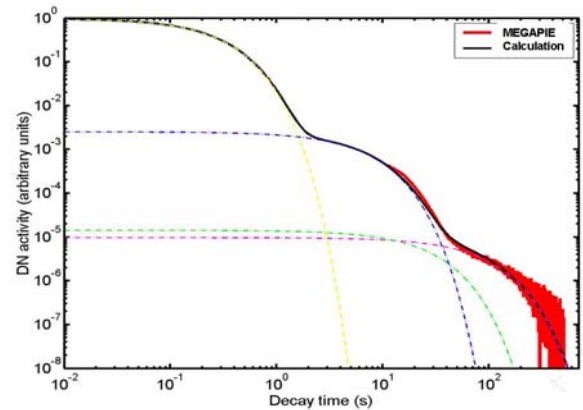

**Fig. 4.** The experimental DN decay curve (in red) compared to the curve (in black) obtained from eq. 1 using the PNPI DN parameters. The relative contributions from individual precursors are also shown (dashed lines: $^{9}$Li in yellow, $^{17}$N in blue, $^{88}$Br in green and $^{87}$Br in purple). Note the arbitrary units in the figure.

## 6 Conclusions and perspectives

We presented a preliminary comparison between the DN decay curve measured at MEGAPIE and the result of a geometrical model involving three averaged liquid metal transit times and the DN precursor parameters. The agreement, using DN parameters previously measured at PNPI, is rather good. A further step will be to fit the experimental curve in order to better estimate the transit times and, fixing the time parameters, to extract DN precursor parameters and compare them to PNPI results.

The authors are grateful to the MEGAPIE project management and PSI-SINQ personnel, in particular to Luca Zanini for cooperation and assistance during the detector installation and data taking.